\documentclass[doublecol,linenumbers]{epl2} 

\usepackage{amstext}
\usepackage{amsmath}
\title{Role of the effective payoff function in evolutionary game dynamics}

\author{Feng Huang\inst{1} \and Xiaojie Chen\inst{2} \and Long Wang\inst{1,*}}

\institute{
  \inst{1} Center for Systems and Control, College of Engineering, Peking University, Beijing 100871, P. R. China\\
  \inst{2} School of Mathematical Sciences, University of Electronic Science and Technology of China, Chengdu 611731, P. R. China \\
  \inst{*} longwang@pku.edu.cn
}

\abstract{
In most studies regarding evolutionary game dynamics, the effective payoff, a quantity that translates the payoff derived from game interactions into reproductive success, is usually assumed to be a specific function of the payoff. Meanwhile, the effect of different function forms of effective payoff on evolutionary dynamics is always left in the basket. With introducing a generalized mapping that the effective payoff of individuals is a non-negative function of two variables on selection intensity and payoff, we study how different effective payoff functions affect evolutionary dynamics in a symmetrical mutation-selection process. For standard two-strategy two-player games, we find that under weak selection the condition for one strategy to dominate the other depends not only on the classical $\sigma$-rule, but also on an extra constant that is determined by the form of the effective payoff function. By changing the sign of the constant, we can alter the direction of strategy selection. Taking the Moran process and pairwise comparison process as specific models in well-mixed populations, we find that different fitness or imitation mappings are equivalent under weak selection. Moreover, the sign of the extra constant determines the direction of one-third law and risk-dominance for sufficiently large populations. This work thus helps to elucidate how the effective payoff function as another fundamental ingredient of evolution affect evolutionary dynamics.}

\begin{document}

\maketitle

\section{Introduction}
In a Darwinian evolutionary process, it mainly incorporates three fundamental ingredients: inheritance, mutation, and selection. Due to the influence of perturbation or random drift, a genetic process frequently accompanies the occurrence of mutations, which creates genotypic or phenotypic variation, thereby finally leading to the differences in individual fitness that selection acts upon~\cite{nowak2006evolutionary}. By integrating game theory with Darwinian evolution~\cite{smith1973logic,Smith1982}, evolutionary game theory has become a powerful mathematical framework to model biological~\cite{turner1999prisoner,kerr2002local} and social~\cite{hamilton1964genetical,nowak2005evolution,helbing2009outbreak,chen2018punishment} evolution in a population consisting of different types of interacting individuals under frequency-dependent selection.
\par
Traditionally, a widely used system that focuses on the effects of frequency-dependent selection is the replicator equation~\cite{taylor1978evolutionary,hofbauer1998evolutionary}, where the population is infinitely large well-mixed and the stochastic effect is exclusively overlooked usually. However, if we relax this setting to a more realistic situation where the population is finite well-mixed and subject to fluctuations, this deterministic approach is augmented and disturbed by random drift~\cite{van1992stochastic,nowak2004emergence,sui2015evolutionary}. In such a finite population with fluctuations, it needs to resort to the tool of stochastic evolutionary game dynamics for investigating the evolution of different traits~\cite{nowak2006evolutionary,taylor2004evolutionary}. In addition to the classical Wright-Fisher process~\cite{van1992stochastic,imhof2006evolutionary}, frequency-dependent Moran process~\cite{nowak2004emergence,taylor2004evolutionary}, and imitation-based pairwise comparison process~\cite{szabo2007evolutionary,altrock2009fixation,wu2010universality} are two most common microscopic models of strategy spreading in finite populations. In contrast to the well-mixed population setting above, there are also lots of interest in studying evolutionary game dynamics in structured populations~\cite{szabo2007evolutionary,szolnoki2008towards,perc2012self,perc2013evolutionary}. Typically, the spatial geometry of population structure is modeled by regular lattices~\cite{nowak1992evolutionary,szabo1998evolutionary,hauert2005game,perc2006chaos,perc2010biosystems,chen2008interaction} or more general complex networks~\cite{ohtsuki2006simple,santos2008social,fu2009evolutionary,chen2008promotion}, where individual interactions merely occur among nearest neighbors.
\par
For a game system, in general, the ingredients influencing the final evolutionary outcomes are nothing but the model, update rule, mutation rate, and population structure, etc. Models and update rules determine the way of strategy spreading. Mutation rates measure the intensity of randomness, while the underlying population structure describes the geometry of individual interactions. Depending on the game interactions, each individual obtains a payoff, and finally it needs to translate into reproductive success, termed effective payoff~\cite{tarnita2009strategy}. For example, the effective payoff is known as fitness in Moran process~\cite{nowak2004emergence,taylor2004evolutionary} and imitation probability in pairwise comparison process~\cite{altrock2009fixation,wu2010universality,huang2018conditional}. Based on the usual assumption that the effective payoff is the form of $1+(\text{Selection intensity})\cdot(\text{Payoff}$), for standard two-strategy two-player games, Tarnita et al. demonstrate that if the selection intensity is weak the condition for one strategy to dominate the other is determined by a `$\sigma$-rule'~\cite{tarnita2009strategy}, which holds on any population structure~\cite{allen2017evolutionary}. The parameter $\sigma$, termed structure coefficient, is a quantity that only depends on the population structure, update rule, and mutation rate, but not on the payoff values. Later, this work attracts wide interest~\cite{tarnita2011multiple,wu2013dynamic,mcavoy2016structure}. For two-player games with multiple strategies, it involves two structure coefficients~\cite{tarnita2011multiple}. To calculate them, investigating games with three strategies is enough. While for multi-player games with two strategies where $d$ individuals are selected to play a game, the $\sigma$-rule will depend on $d-1$ structure coefficients~\cite{wu2013dynamic}. In particular, for a more general setting of multi-player games with many strategies, this rule turns out to be quite complicated and the number of structure coefficients required for a symmetric game with $d$-player and $n$-strategy grows in $d$ like $d^{n-1}$~\cite{mcavoy2016structure}. Clearly, because the form of the effective payoff is a specific function in these works, the $\sigma$-rule does not reflect the influence of the effective payoff function on evolutionary outcomes.
\par
In evolutionary biology, however, how to measure the genotype-fitness map (i.e., the fitness landscape) is always a challenging issue, and now it has been accepted that the shape of the genotype-fitness map has fundamental effects on the course of evolution~\cite{de2014empirical}. In addition, based on a Markov chain model, it has been demonstrated that the heterogeneity of individual background fitness can act as a suppressor of selection~\cite{hauser2014heterogeneity}. In a way, therefore, it means that the form of the effective payoff function (which translates the payoff derived from game interactions into the ability of reproductive success) has a significant effect on the evolution of game dynamics.
\par
To this end, in this letter we study the effect of different function forms of effective payoff on evolutionary dynamics and accordingly extend the results given by Tarnita et al.~\cite{tarnita2009strategy}. We find that if the first-order derivative of the effective payoff function can be written by a linear combination of payoff, then the condition for one strategy to dominate the other depends not only on the $\sigma$-rule, but also on an extra constant which is determined by the effective payoff function. This constant determines the direction of $\sigma$-rule (strategy selection). Additionally, taking the Moran process and pairwise comparison process as specific models in well-mixed populations, we demonstrate that different fitness or imitation mappings are equivalent under weak selection and the extra constant curbs the direction of one-third law and risk-dominance in the limit of large populations.

\section{\label{s1}Model and results}
In a structured population with $N$ individuals, we consider stochastic evolutionary dynamics induced by a mutation-selection process. Each player can choose an arbitrary strategy from $A$ and $B$. Then, depending on the payoff matrix
\begin{equation}\label{eq1}
\bordermatrix{
~ &A  &B    \cr
A& a & b             \cr
B & c & d                 \cr
},
\end{equation}
players obtain an accumulative payoff by interacting with other individuals based on the underlying population structure. For example, when an $A$ player interacts with another $A$ player, it will obtain a payoff $a$, but $b$ when interacting with a $B$ player. Likewise, a $B$ player can obtain a payoff $c$ when interacting with an $A$ player, and payoff $d$ when interacting with another $B$ player. Therefore, the total payoff of each player is a linear function of $a$, $b$, $c$, and $d$ without including constant terms (evidently, if the payoff is calculated in an average way, the linear relation also holds). For an $A$ player, for instance, the total payoff is $a\cdot(\text{number of \textit{A}-neighbors})+b\cdot(\text{number of \textit{B}-neighbors})$. To study the effect of effective payoff functions on evolutionary dynamics, instead of a specific form, we assume that the effective payoff of a player is given by $\varphi(\beta,\text{Payoff})$. Parameter $\beta$ measures the intensity of selection, and $\beta\rightarrow0$ corresponds to the case of weak selection~\cite{nowak2004emergence,wu2010universality,wu2013extrapolating}.
\par
The reproductive process of each player is dependent on the update rule and its effective payoff, and subject to mutations. With probability $\mu$, a mutation occurs and the offspring adopts a strategy ($A$ or $B$) at random. Otherwise, with probability $1-\mu$, the offspring inherits its parent's strategy. For $\mu=1$, there are only mutations, no selection, and strategy choice is completely random. If $0<\mu<1$, however, there exists a mutation-selection equilibrium~\cite{fudenberg2006evolutionary,antal2009strategy}.
\par
For the game of two strategies, the frequency of $A$ players in the population defines a finite state space, $S$, and the evolutionary dynamics can be captured by a Markov process on this state space. We denote the transition probability from state $S_i$ to state $S_j$ by $P_{ij}$. Since the transition probability depends on the update rule and on the effective payoff of players, it can be given by $P_{ij}[\varphi(\beta,\text{Payoff})]$. Furthermore, we assume that $\varphi(\beta,\text{Payoff})$ is differentiable at $\beta=0$. In the limit of weak selection, then we can give $\varphi(\beta,\text{Payoff})$ in the form of first-order Taylor expansions,
$\varphi(\beta,\text{Payoff})=\varphi_0+\varphi^{(1)}(0)
  \cdot \beta+o(\beta)$.
Here $\varphi_0:=\varphi(0,\text{Payoff})$ represents the baseline effective payoff of each player, and $\varphi^{(1)}(0):=[\partial \varphi(\beta,\text{Payoff})/ \partial \beta]_{\beta=0}$ represents the first-order coefficient of selection intensity. Particularly, if $\varphi^{(1)}(0)$ can be written by a linear combination of payoff, that is $\varphi^{(1)}(0)=[\partial \varphi(\beta,\text{Payoff})/ \partial \beta]_{\beta=0}=k_0\cdot \text{Payoff}+c_0$, then the transition probability is given by $P_{ij}[\varphi_0+(k_0\cdot \text{Payoff}+c_0)\beta+o(\beta)]$. Clearly, the constants, $k_0$ and $c_0$, are dependent on the choice of the function $\varphi(\beta,\text{Payoff})$ and may rely on the entries of the payoff matrix. In addition, actually a large body of functions meet the condition $\varphi^{(1)}(0)=k_0\cdot \text{Payoff}+c_0$, such as $1-\beta+\beta\cdot\text{Payoff}$~\cite{nowak2004emergence}, $exp(\beta\cdot\text{Payoff})$~\cite{wu2013dynamic}, and $\varphi(\beta\cdot\text{Payoff})$~\cite{wu2010universality}. Therefore, this condition is not a harsh requirement. Note that the payoff of players is linear in $a$, $b$, $c$, and $d$ without constant terms, it follows that the transition probability is the function $P_{ij}[(k_0a+c_0)\beta,(k_0b+c_0)\beta,(k_0c+c_0)\beta,(k_0d+c_0)\beta]$. Then, based on the notation $(a',b',c',d'):=(k_0a+c_0,k_0b+c_0,k_0c+c_0,k_0d+c_0)$ and following the proof given in Ref.~\cite{tarnita2009strategy}, we know that the condition that strategy $A$ is favored over strategy $B$ (i.e., strategy $A$ is more abundant than $B$ in the mutation-selection equilibrium) is $\sigma a'+b'>c'+\sigma d'$, where $\sigma$ is a parameter that depends on the population structure, update rule, and mutation rate. Accordingly, we have the following theorem:
\\
\noindent\textbf{Theorem $1$.}  Consider a population structure and an update rule that satisfy the following three conditions: (i) the transition probabilities are differentiable at $\beta=0$; (ii) the update rule is symmetric for the two strategies $A$ and $B$; and (iii) in the game given by the matrix entries, $a=c=d=0$ and $b=1$, strategy $A$ is not disfavored. Then, in the limit of weak selection, when the effective payoff function $\varphi(\beta,\text{Payoff})$ satisfies $[\partial \varphi(\beta,\text{Payoff})/ \partial \beta]_{\beta=0}=k_0\cdot \text{Payoff}+c_0$, strategy $A$ is favored over strategy $B$ if $
  k_0(\sigma a+b-c-\sigma d)>0,
$
where $k_0$ is a constant that relies on the function of the effective payoff, and $\sigma$ is the structure coefficient which depends on the model and the dynamics (population structure, update rule, and mutation rate), but not on the entries of the payoff matrix.
\par
This theorem implies that for determining the condition under which one strategy dominates the other, the classical $\sigma$-rule~\cite{tarnita2009strategy} is not enough. It also depends on an additional constant $k_0$ determined by the effective payoff function. Actually, the constant $k_0$ controls the direction of $\sigma$-rule (strategy selection). If the effective payoff function is given such that the constant $k_0$ is positive, the theorem recovers the classical $\sigma$-rule (selection favors $A$ to dominate $B$). Otherwise, if the effective payoff function is given such that $k_0$ is negative, the classical $\sigma$-rule will reverse the direction (selection favors $B$ to dominate $A$).

\section{\label{s2}Moran and pairwise comparison process}
To check the validity of our theorem and to study how the effective payoff function influences the evolutionary outcomes in a specific dynamic process, here we consider the frequency-dependent Moran process and pairwise comparison process. These two processes represent two classes of typical evolutionary dynamics. The former describes how successful strategies spread in the population through genetic reproduction, whereas the latter describes such a process through cultural imitation.
\par
In the Moran process, the effective payoff is known as the individual fitness, which measures the ability to survive and produce offspring. With a probability proportional to the fitness, an individual is selected randomly for reproduction. And then one identical offspring replaces another randomly chosen individual. Usually, the fitness is assumed to be a convex combination of a background fitness (which is set to one) and the payoff from the game~\cite{nowak2004emergence,taylor2004evolutionary}, or an exponential function of payoff~\cite{wu2013dynamic}. Under these specific forms, the constant $k_0$ related to the fitness function actually turns to $1$ and it reduces to the previous results~\cite{nowak2004emergence,altrock2009fixation}. Instead of using a specific fitness form, here we adopt a generalized mapping that the fitness of a player is any a non-negative function of two variables on selection intensity and payoff, $f(\beta,\text{Payoff})$. Since mutation occurs during the process of reproduction, it follows that the transition probabilities are given by
\begin{equation}\label{eq4}
\begin{split}
  P_{i,i+1}&=\frac{if(\beta,\pi_A)(1-\mu)+(N-i)f(\beta,\pi_B)\mu}
  {if(\beta,\pi_A)+(N-i)f(\beta,\pi_B)}\frac{N-i}{N}, \\
   P_{i,i-1}&=\frac{(N-i)f(\beta,\pi_B)(1-\mu)+if(\beta,\pi_A)\mu}
  {if(\beta,\pi_A)+(N-i)f(\beta,\pi_B)}\frac{i}{N},
\end{split}
\end{equation}
where $\pi_A(i):=[a(i-1)+b(N-i)]/(N-1)$ and $\pi_B(i):=[ci+d(N-i-1)]/(N-1)$ are the average payoffs of an $A$ player and a $B$ player, respectively.
\par
While in the pairwise comparison process, the effective payoff is known as the imitation probability. Two individuals are sampled randomly and then a focal player imitates the strategy of the role model with a probability depending on the payoff comparison~\cite{antal2009strategy,altrock2009fixation}. As usual, the imitation probability is modeled by the Fermi function with considering the
effect of noise~\cite{szabo1998evolutionary,szabo2007evolutionary,
hauert2005game}. Thus, this process is also called Fermi process. Under the situation that the effective payoff mapping is non-specified, however, the imitation probability function for the pairwise comparison process should be given by $g(\beta,\Delta\pi)$. Here, $\Delta\pi$ denotes the difference of average payoffs between strategy $A$ and $B$. In the presence of mutations, this imitation process occurs accurately with probability $1-\mu$. Otherwise, with probability $\mu$, the focal player adopts a random strategy, $A$ or $B$. Then, it leads to the transition probabilities,
\begin{equation}\label{eq5}
\begin{split}
  P_{i,i+1}&=(1-\mu)\frac{i}{N}\frac{N-i}{N}g(\beta,\Delta\pi(i))
  +\frac{N-i}{N}\frac{\mu}{2}, \\
  P_{i,i-1}&=(1-\mu)\frac{i}{N}\frac{N-i}{N}g(\beta,-\Delta\pi(i))
  +\frac{i}{N}\frac{\mu}{2},
\end{split}
\end{equation}
where $\Delta\pi(i):=\pi_{A}(i)-\pi_{B}(i)=ui+v$. Herein, parameters $u$ and $v$ are defined by $u:=(a-b-c+d)/(N-1)$ and $v:=(Nb-Nd-a+d)/(N-1)$, respectively.
\par
Moreover, for both processes, the probability to stay in the current state is $1-P_{i,i+1}-P_{i,i-1}$, and the probability to transform to other states is vanishing. In what follows, we first calculate the fixation probabilities and fixation times under weak selection when mutations are absent. Then, we derive the criterion that strategy $A$ is favored over strategy $B$ in this case, and finally extrapolate this criterion to small mutation rates.

\subsection{Fixation probabilities and fixation times}
If there are no mutations in these two game systems, then one quantity of most interest is the fixation probability, $\phi_i$, which describes the probability that $i$ individuals of type $A$ reach fixation at all $A$. Another significant quantity is the average time for a single $A$ player reaching fixation~\cite{antal2006fixation,taylor2006symmetry,sui2015speed}.
The former measures the preference of natural selection whereas the latter characterizes the evolutionary velocity of the system.
\par
First, we follow the conditions given by \textbf{Theorem $1$}, that is, the first-order derivative of fitness function for the Moran process can be written by $f_{\beta}(0,\pi)=k_0^{(m)}\cdot\pi+c_0^{(m)}$, and the one of imitation probability function for the pairwise comparison process can be written by $g_{\beta}(0,\Delta\pi)=k_0^{(p)}\cdot\Delta\pi+c_0^{(p)}$. Here, $k_0^{(m)}$ and $c_0^{(m)}$ ($k_0^{(p)}$ and $c_0^{(p)}$) are two constants that depend on the choice of fitness (imitation probability) functions and may be related to the entries of the payoff matrix. With the notations $m_0:=k_0^{(m)}/f_0$ and $p_0:=2k_0^{(p)}/g_0=k_0^{(p)}/g_{0}^{2}$, where $f_0=f(0,\pi)$ is the baseline fitness of each player and $g_0=g(0,\Delta\pi)=1/2$ is the probability of random imitation, we obtain the approximation of fixation probabilities under weak selection (details for the Supplementary Material) as
\begin{equation}\label{eq5a}
  \phi_i\approx\frac{i}{N}+\beta\cdot s\cdot\frac{i(N-i)[(N+i)u+3v]}{6N},
\end{equation}
where $s=m_0$ for the frequency-dependent Moran process, and $s=p_0$ for the pairwise comparison process.
\par
While for the average times of a single $A$ player reaching fixation, there are two kinds of fixation times that attract much research attention~\cite{antal2006fixation,altrock2009fixation}.
The first one is the unconditional average time of fixation $t_1$, which is the expected value for the time until the population reaches one of the two absorbing states, all $A$ and all $B$, when starting from a single $A$. Another is the conditional average time of fixation $t_1^A$, which specifies the expected time that a player of type $A$ takes to reach the absorbing state, all $A$. In the limit of weak selection, we find that the unconditional and conditional fixation times for the Moran process (details for the Supplementary Material) can be approximated to
\begin{equation}\label{eq7}
  \begin{split}
    t_1&\approx
    NH_{N-1}+m_0v\frac{N}{2}(N+1-2H_N)\beta, \\
    t_1^A&\approx
    N(N-1)-m_0u\frac{N^2(N^2-3N+2)}{36}\beta,
  \end{split}
\end{equation}
whereas for the pairwise comparison process, they are given by
\begin{equation}\label{eq8}
\begin{split}
t_1\approx 2NH_{N-1}+p_0vN(N-1-H_{N-1})\beta,  \\
t_1^A\approx 2N(N-1)-p_0uN(N-1)\frac{N^2+N-6}{18}\beta,
\end{split}
\end{equation}
where $H_{N}=\sum_{l=1}^{N}\frac{1}{l}$ is the harmonic number.
\par

\begin{figure}
\includegraphics[width=\hsize]{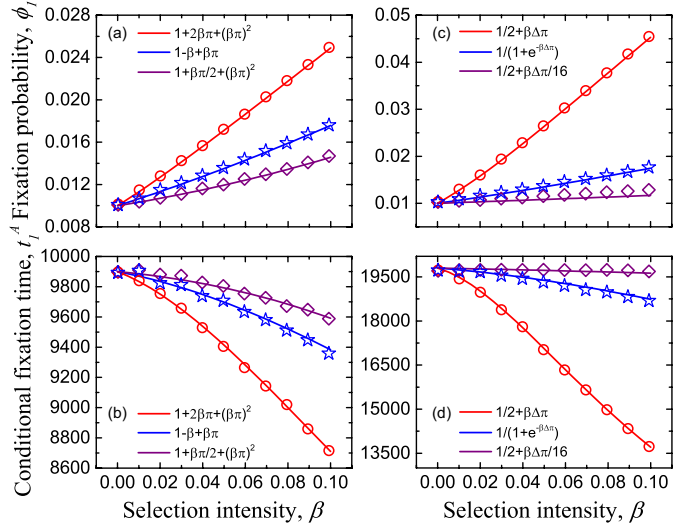}
\caption{\label{Fig0} Effect of fitness and imitation probability functions on evolutionary dynamics acts as a selection amplifier or suppressor. Using the payoff matrix entries, $a=1.2$, $b=0.8$, $c=1.0$, and $d=0.7$ (i.e., $A$ dominates $B$), we show the fixation probabilities and conditional fixation times of a single $A$ when population size is $N=100$ in Moran process (panel (a) and (b)) and pairwise comparison process (panel (c) and (d)), respectively. Lines are analytical results (Eqs.~(S$9$) and (S$14$) in the Supplementary Material), while symbols are simulations. If a fitness function is applied such that $m_0>1$ ((a) and (b)) or an imitation probability function such that $p_0>1$ ((c) and (d)), it promotes the fixation of advantage strategy and decreases the fixation time (red lines) compared with the benchmark $m_0=1$ or $p_0=1$ (blue lines). Otherwise, if $m_0<1$ or $p_0<1$, it suppresses the fixation of advantage strategy and increases the fixation time (purple lines) compared with the benchmark (blue lines).}
\end{figure}

Interestingly, for both Moran process and pairwise comparison process, if the first-order derivative of the effective payoff function (i.e., the fitness and imitation probability function) can be written by a linear combination of the payoff, the difference in the influence of effective payoff functions on evolutionary outcomes just embodies in the coefficients before selection intensity, $m_0$ and $p_0$. By proper rescaling, actually, these constant coefficients can be absorbed into the selection intensity, or make all payoff matrix entries ($a$, $b$, $c$, and $d$) change a scale in view of the exact formulae of $u$ and $v$. In particular, if we adopt a linear or an exponential form of payoff as the fitness function, or the Fermi function as the imitation probability, both $m_0$ and $p_0$ are $1$, which recovers the previous results~\cite{altrock2009fixation} as specific cases. Moreover, under the conditions $[(N+1)u+3v]>0$ and $u>0$, if a fitness (imitation probability) function is chosen such that $m_0>0$ ($p_0>0$), then taking the constant $1$ as the benchmark, $m_0>1$ ($p_0>1$) leads this function to acting as an amplifier of selection (facilitating the fixation of advantage individuals and decreasing the fixation time). Nevertheless, when $m_0<1$ ($p_0<1$), this function acts as a suppressor of selection (suppressing the fixation of advantage individuals and increasing the fixation time). This result holds not only for weak selection, but also for intermediate selection intensity (see Fig.~\ref{Fig0}). As the counterpart, if the function is chosen such that $m_0<0$ ($p_0<0$), with the scaling theory of dilemma strength~\cite{wang2015dilemma,tanimoto2007relationship}, another example of prisoner's dilemma is given in the Supplementary Material.

\subsection{Equivalence}
Based on the above calculations of fixation probabilities and fixation times under weak selection, additionally, we find that two arbitrary fitness (imitation probability) functions in a Moran process (pairwise comparison process) are equivalent. Specifically, for the frequency-dependent Moran process with a generalized fitness function $f(\beta,\pi)$, if $f_{\beta}(0,\pi)=k_0^{(m)}\cdot\pi+c_0^{(m)}$ is satisfied, we know that the influence of any two different fitness functions on evolutionary outcomes just embodies in the constant factor $m_0$ before the selection intensity. Thus, in this sense, any two fitness mappings meeting the conditions defined above are equivalent under weak selection. The equivalence means that the difference in fixation probabilities and fixation times is merely captured by the constant factor $m_0$, with which the payoff matrix or the intensity of selection changes a scale.
\par
\begin{figure}
\centering
\includegraphics[width=\hsize]{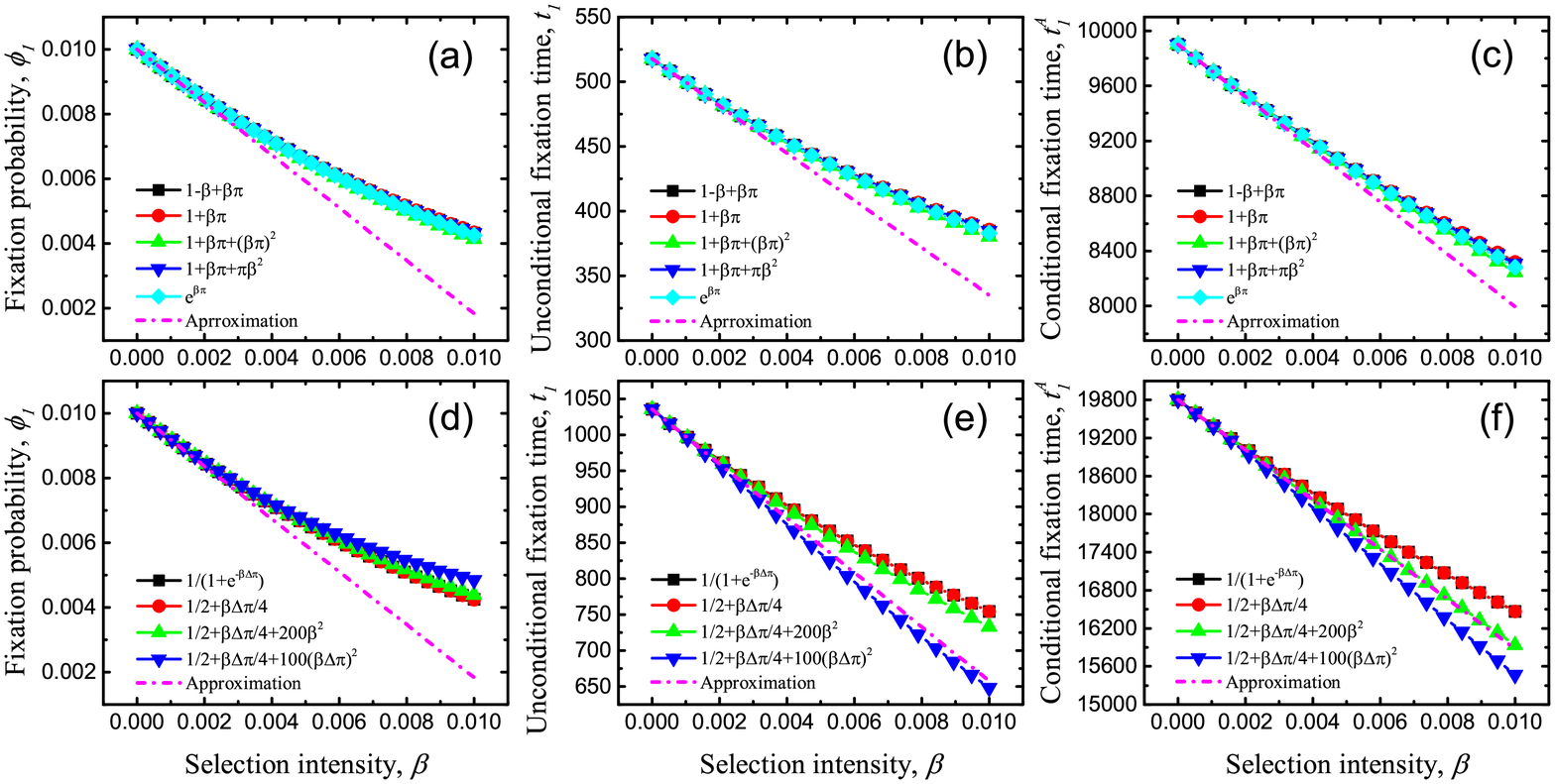}
\caption{\label{fig1} Equivalence of Moran process and pairwise comparison process under weak selection. In the first row, we show the fixation probability $\phi_1$ (a), unconditional fixation time $t_1$ (b), and conditional fixation time $t_1^A$ (c) of a single mutant $A$ in Moran process, respectively. The exact analytical results are depicted by solid lines with symbols, and accordingly we show the weak selection approximations by short dash dots, based on a class of equivalent fitness functions. The same manipulation is repeated for pairwise comparison process with a class of equivalent imitation probability functions, as shown in the second row. Analytical results are numerical calculations on the basis of exact Eqs.~(S$9$), (S$12$), and (S$14$) in the Supplementary Material, but weak selection approximations are based on Eqs.~(\ref{eq5a})--(\ref{eq8}). Parameters are $N=100$, $a=4$, $b=1$, $c=1$, and $d=5$ (coordination games) in all panels. (These results are also valid for dominance and coexistence games.)}
\end{figure}

Particularly, if the fitness function adopts one of the function families, $F_1(\beta,\pi)=\sum_{j=0}^{m}a_j\beta^j+
\sum_{i=1}^{n}b_i(\beta\pi)^i$ and $F_2(\beta,\pi)=\sum_{j=0}^{m}a_j\beta^j
+\sum_{i=1}^{n}b_i\beta^i\pi\ (m,n=1,2,3,...)$, where $a_j$ and $b_i$ are constant coefficients that guarantee $F_1>0$ and $F_2>0$ because the individual fitness is positive~\cite{wu2013dynamic,wu2010universality}, then we have the same factor $m_0=b_1/a_0$. Interestingly, if $m_0=b_1/a_0=1$, these two function families are equivalent to the prevalent fitness mappings $1-\beta+\beta\pi$ and $exp(\beta\pi)$ under weak selection (see Fig.~\ref{fig1}). Actually, the Taylor series of $exp(\beta\pi)$ at $\beta=0$ is just the function family $F_1$ when specific coefficients $a_0=1$, $a_j=0\ (j>0)$, and $b_i=1/(i!)$ are applied.
\par
Similarly, for the pairwise comparison process with a generalized imitation probability function $g(\beta,\Delta\pi)$, if  $g_{\beta}(0,\Delta\pi)=k_0^{(p)}\cdot\Delta\pi+c_0^{(p)}$ is satisfied, then the influence of any two different imitation probability functions on evolutionary outcomes also embodies in a constant factor before the selection intensity, $p_0$. Thus, in this sense, for two arbitrary imitation probability functions, they are also equivalent under weak selection. The equivalence follows the same meaning as that in the Moran process, that is, the constant factor $p_0$ uniquely measures the difference of fixation probabilities and fixation times under weak selection.
\par
Surprisingly, with a completely similar formulation to the fitness function families $F_1$ and $F_2$, $G_1(\beta,\Delta\pi)=\sum_{j=0}^{m}\alpha_j\beta^j+
\sum_{i=1}^{n}\eta_i(\beta\Delta\pi)^{i}$ and $G_2(\beta,\Delta\pi)=\sum_{j=0}^{m}\alpha_j\beta^j+\sum_{i=1}^{n}
\eta_i\beta^i\Delta\pi\ (m,n=1,2,3,...)$, where $\alpha_j$ and $\eta_i$ are constant coefficients which guarantee that $G_1$ and $G_2$ are probability functions and $\alpha_0=1/2$, are two classes of equivalent imitation probability functions. Particularly, if $\eta_1=1/4$, these two function families are equivalent to the popular Fermi function $1/(1+e^{-\beta\Delta\pi})$ under weak selection (see Fig.~\ref{fig1}). Actually, by choosing specific coefficients, $G_1$ can also become the Taylor series of Fermi function at $\beta=0$.

\subsection{One-third law and risk-dominance}
\begin{figure}
\includegraphics[width=\hsize]{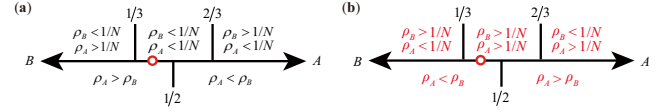}
\caption{\label{Fig2} Direction of the one-third law and the risk dominance is determined by the choice of fitness and imitation probability functions. When $k_0^{(m)}>0$ for Moran process or $k_0^{(p)}>0$ for pairwise comparison process, it corresponds to the classical one-third law and risk-dominance in coordination games (a). While if $k_0^{(m)}<0$ for Moran process or $k_0^{(p)}<0$ for pairwise comparison process, the direction of one-third law and risk-dominance is reversed (b). Otherwise, the conditions determining $\rho_A>1/N$ or $\rho_A>\rho_B$ under weak selection will depend on higher order coefficients of fixation probabilities.}
\end{figure}

In stochastic evolutionary game dynamics, the notions of invasion and fixation are two fundamental concepts to describe the spreading of strategies in finite populations~\cite{nowak2004emergence,taylor2004evolutionary}. Using the neutral game as the benchmark, strategy $A$ is shortly said to fixate in a resident population (selection favors $A$ replacing $B$) if the fixation probability for a single $A$ is larger than that in the neutral game~\cite{nowak2004emergence,taylor2004evolutionary}. Thus, for the frequency-dependent Moran process with a fitness function $f(\beta,\pi)$ which fulfills the condition given above, selection favors $A$ replacing $B$ under weak selection if $m_0[(N+1)u+3v]>0$ (see Eq.~(\ref{eq5a})). Similarly, for the pairwise comparison process with an imitation probability function $g(\beta,\Delta\pi)$, selection favors $A$ replacing $B$ under weak selection if $p_0[(N+1)u+3v]>0$ (see Eq.~(\ref{eq5a})). In view of the notations $m_0:=k_0^{(m)}/f_0$ and $p_0:=2k_0^{(p)}/g_0=k_0^{(p)}/g_{0}^{2}$, it follows that the criterion that selection favors $A$ replacing $B$ under weak selection is $k_0[(N+1)u+3v]>0$, where $k_0=k_0^{(m)}$ for the Moran process and $k_0=k_0^{(p)}$ for the pairwise comparison process. Specifically, we have (see Fig.~\ref{Fig2}): (i) When $k_0>0$, the condition under which selection favors $A$ replacing $B$ ($\phi_1>1/N$) is $a(N-2)+b(2N-1)>c(N+1)+d(2N-4)$. Particularly, for sufficiently large population size $N\rightarrow+\infty$, it corresponds to one-third law~\cite{nowak2004emergence} in the case of coordination games ($1/3>(d-b)/(a-b-c+d)$); (ii) When $k_0<0$, this condition changes to $a(N-2)+b(2N-1)<c(N+1)+d(2N-4)$. Particularly, for $N\rightarrow+\infty$, the classical one-third law is reversed in the case of coordination games ($1/3<(d-b)/(a-b-c+d)$); (iii) When $k_0=0$, the condition that selection favors $A$ replacing $B$ will depend on the higher order coefficients of $\beta$ in $\phi_1$. Actually, the calculations of higher order coefficients under weak selection are more tedious than the linear approximation~\cite{wu2010universality}.
\par
Except for the underlying principle that determines the condition of favoring strategy $A$ to replace $B$, it is also of interest to ask whether strategy $A$ is selected over strategy $B$, termed `strategy selection'~\cite{tarnita2009strategy}. First, let $\rho_A$ ($\rho_B$) denotes the fixation probability that a single individual using strategy $A$ ($B$) invades and takes over a resident population of $B$ ($A$) players. Accordingly, we have $\rho_A=\phi_1$. Moreover, note that the probability $\rho_B$ is equal to that $N-1$ individuals of type $A$ fail to take over a population in which there is just a single $B$ individual. Then, it leads to $\rho_B=1-\phi_{N-1}$. With introducing the notation $\gamma_j=P_{j,j-1}/P_{j,j+1}$, which is the ratio of transition probabilities when mutations are absent ($\mu=0$), we have the ratio of fixation probabilities for strategy $A$ and $B$ as $\rho_B/\rho_A=\prod_{j=1}^{N-1}\gamma_j$. Under weak selection, the ratio of these two fixation probabilities can be approximated to
\begin{equation}
\begin{split}
  \rho_B/\rho_A &\approx 1-s\beta\sum_{j=1}^{N-1}(uj+v) \\
  &=1-\frac{s\beta}{2}[a(N-2)+bN-cN-d(N-2)],
\end{split}
\end{equation}
where $s=m_0$ for the Moran process, and $s=p_0$ for the pairwise comparison process. In view of the definitions of $m_0$ and $p_0$, therefore, the condition under which strategy $A$ is selected over strategy $B$ is given by $k_0(\sigma a+b-c-\sigma d)>0$, where $k_0=k_0^{(m)}$ for the Moran process and $k_0=k_0^{(p)}$ for the pairwise comparison process, and $\sigma=(N-2)/N$ is the structure coefficient of well-mixed populations~\cite{tarnita2009strategy}. Under weak selection, specifically we have (see Fig.~\ref{Fig2}): (i) When $k_0>0$, the condition that strategy $A$ is selected over strategy $B$ is $\sigma a+b>c+\sigma d$. Particularly, for sufficiently large population size $N\rightarrow+\infty$, it corresponds to that $A$ is risk-dominant in the case of coordination games ($1/2>(d-b)/(a-b-c+d)$); (ii) When $k_0<0$, this condition changes to $\sigma a+b<c+\sigma d$. Particularly, for $N\rightarrow+\infty$, it corresponds to that $B$ is risk-dominant in the case of coordination games ($1/2<(d-b)/(a-b-c+d)$); (iii) When $k_0=0$, the condition that strategy $A$ is selected over strategy $B$ under weak selection will depend on higher order coefficients of $\beta$ in $\rho_B/\rho_A$.
\par
In particular, if we additionally consider the situation where small non-uniform mutations occur between the two strategies, that strategy $A$ is more abundant than $B$ in the long run is determined by
$
  \mu_{AB}\rho_B/\mu_{BA}\rho_A<1
$~\cite{fudenberg2006evolutionary,antal2006fixation,wu2012small},
where $\mu_{AB}$ and $\mu_{BA}$ denote the mutation rates from $A$ to $B$ and from $B$ to $A$, respectively. For $\mu_{AB}=\mu_{BA}$, clearly the conclusions obtained above are still valid, which extrapolates our results to the situation of small mutations.

\section{\label{s5}Discussion}
Most previous studies exclusively assume that the effective payoff function is a specific form. One direct result of this setting is that the final evolutionary outcomes do not reflect the effect of the effective payoff function on evolutionary dynamics. Therefore, it still remains unclear how the effective payoff function influences the evolutionary dynamics in a game system. With introducing a generalized mapping, we investigate such effect in this letter.
\par
For standard $2\times2$ games where a specific form of the effective payoff is adopted, it has been demonstrated that under weak selection the condition for one strategy to dominate the other is determined by a $\sigma$-rule~\cite{tarnita2009strategy}. This rule almost captures all aspects of evolutionary dynamics, but ignores the effect of the effective payoff function. Particularly, if the effective payoff function is an any non-negative mapping of the product of the payoff and the selection intensity~\cite{wu2010universality,wu2013dynamic}, the rule still holds. But it does not change the basic fact that the role played by the effective payoff function in evolutionary game dynamics is exclusively overlooked. With introducing a more generalized mapping that the effective payoff of individuals is a function of payoff and selection intensity, however, we find that the condition determining a strategy to be selected relies not only on the $\sigma$-rule, but also on an extra constant which characterizes and depends on the effective payoff function. As an extension, a multi-strategy version is also given in the Supplementary Material.
\par
Based on specific effective payoff forms, it has been found that the linear and exponential functions lead to identical evolutionary outcomes under weak selection~\cite{antal2009strategy,wu2013dynamic,wu2010universality}. Here we generalize this equivalence understanding to any two fitness functions in a Moran process, and imitation probability functions in a pairwise comparison process. In addition, except for the standard $2\times2$ games and weak selection, there are lots of research interest in the games of multiple players or strategies~\cite{wu2013dynamic,su2018understanding,tarnita2011multiple} and strong selection~\cite{altrock2017evolutionary}, which are worth the effort in the future.

\acknowledgments
We thank Bin Wu for helpful discussions and comments. This work was
supported by the National Natural Science Foundation of China (Grants No. 61751301, No. 61533001, and No.
61503062).

\nocite{*}
\bibliography{apssamp}

\end{document}